\documentstyle[aps,epsfig,prb]{revtex}  			%twocolumn
\begin{document} 
\twocolumn[\hsize\textwidth\columnwidth\hsize\csname  	 
%twocolumn
@twocolumnfalse\endcsname				%twocolumn 
\draft
\title{Fermi-Liquid Interactions in $d$-Wave Superconductors}
\author{M. B. Walker}
\address{Department of Physics, 
University of Toronto,
Toronto, Ont. M5S 1A7 }
\date{\today }
\maketitle

\widetext					%
\begin{abstract}
This article develops a quantitative quasiparticle model of
the low-temperature properties of $d$-wave superconductors which
incorporates both Fermi-liquid effects and
band-structure effects.  The Fermi-liquid 
interaction effects are found to be 
classifiable into strong and negligible renormalizaton
effects, for symmetric and antisymmetric combinations of the 
energies of $k\uparrow$ and $-k\downarrow$ quasiparticles, respectively.  
A particularly
important conclusion is that the leading clean-limit 
temperature-dependent correction to
the superfluid density is not renormalized by Fermi-liquid interactions,
but is subject to a Fermi velocity (or mass) renormalization effect.
This
leads to difficulties in accounting for the penetration depth 
measurements with physically acceptable parameters, and hence reopens
the question of the quantitative validity of the quasiparticle picture.
\end{abstract}


\vfill		%twocolumn
\narrowtext			%

\vskip2pc]	%twocolumn

%\Newpage
There is now considerable experimental evidence
that the  cuprate high $T_c$ superconductors exhibit the 
simple power law temperature dependences predicted by the
quasiparticle picture for their thermodynamic 
and transport properties
at temperatures well below $T_c$.
For example,
penetration depth measurements find that the superfluid
density exhibits a low-temperature clean-limit linear 
in T temperature dependence \protect\cite{har93},
in agreement with theory \protect\cite{pro91}. The NMR
relaxation rate exhibits the expected $T^3$ temperature
dependence \protect\cite{mar93}.  The predicted effect of
impurities in giving rise to a universal thermal conductivity
\protect\cite{lee93,gra96} has been confirmed \protect\cite{tai97}.
The clean limit specific heat varying as $T^2$ appears to have
been observed \protect\cite{mol97,wri99}.
Even the electrical transport relaxation rate observed in
microwave conductivity experiments \protect\cite{hos99}, which
had resisted explanation for some time, has  now been explained
in terms of a quasiparticle picture \protect\cite{wal00}.

Whether or not the magnitudes of the coefficients 
of the above power law temperature dependences are 
accurately given by a quasiparticle description is at
present an open question.  A recent study correlating these
different coefficients \protect\cite{chi00} concludes that the 
quasiparticle model may be successful here also
provided a Fermi-liquid interaction
factor multiplying the superfluid density is treated as an
adjustable parameter.  The contention of this article is that
there is however  no Fermi-liquid interaction renormalization of
the linear in T contribution to the inverse square 
penetration depth.  There is instead a renormalization by a
factor involving the ratio of a band Fermi velocity to a
Landau quasiparticle Fermi velocity.  The difficulty now is
that a physically unreasonable value of this renormalization
parameter is obtained from experiment.
This reopens the question of to what extent the quasiparticle
picture can provide an accurate quantitative picture of the
low temperature behavior of the high $T_c$ superconductors.
Recent debate on correctness of the quasiparticle picture
is also occurring in connection with ARPES experiments
\protect\cite{kam00,val99}, and
in connection with the role of phase fluctuations of the
complex order parameter 
in the determination of the temperature
dependence of the superfluid density \protect\cite{car99}. 

The potential importance of Fermi-liquid interactions in 
renormalizing the superfluid density has been emphasized
in Refs.\ \protect\onlinecite{mil98} and  
\protect\onlinecite{dur00}.
These papers note that Fermi-liquid renormalization effects in 
$d$-wave superconductors can be either strong or weak
according as the contributing quasiparticles are from
the entire Fermi surface, or confined the the nodal
points where the $d$-wave gap goes to zero.
Both expect a
strong renormalization effect for the 
superfluid density, whereas this article does not find 
such an effect.  

This article shows that the physics of Fermi liquid
effects in $d$-wave superconductors has an interesting 
symmetry property.
This manifests itself when the quasiparticle energies
are separated into parts that are ${\mathcal{S}}$ymmetric and 
${\mathcal{A}}$ntisymmetric combinations of the energies of the
$+k\uparrow$ and $-k\downarrow$ states. (The calligraphic letters
${\mathcal{S}}$ and ${\mathcal{A}}$ are used here to emphasize the
difference with the more usual definition of the symmetric and 
antisymmetric combinations with respect to $+k\uparrow$ and
$+k\downarrow$ states common in normal state analyses,
e.g. see Eq. 1.32 of Ref.\ \protect\onlinecite{pin89}.) 
In the presence of
Fermi-liquid interactions, the ${\mathcal{S}}$ymmetric and 
${\mathcal{A}}$ntisymmetric corrections to the quasiparticle
energies obey integral equations that are independent of each other,
and they are renormalized differently.  The 
${\mathcal{S}}$ymmetric 
energy corrections exhibit strong Fermi-liquid renormalization effects, 
while the the ${\mathcal{A}}$ntisymmetric energy corrections 
exhibit relatively weak temperature-dependent renormalizations that
can often be neglected.

Temperature gives a ${\mathcal{S}}$ymmetric correction to the
quasiparticle energy because $+k\uparrow$ and $-k\downarrow$ states
are affected in the same way by temperature.  A superfluid flow
generates an ${\mathcal{A}}$ntisymmetric correction since the
components of $+k$ and $-k$ along the superfluid velocity have
opposite signs.  Also the Zeeman interaction generates an 
${\mathcal{A}}$ntisymmetric correction because the spin $\uparrow$
and spin $\downarrow$ contributions to the energy have opposite 
signs.  Thus the superfluid density and the magnetic susceptibility
are negligibly renormalized by Fermi-liquid interactions, while
the effects of temperature (although relatively small) are strongly
renormalized by Fermi-liquid interactions.

The approach of this article to the inclusion of Fermi-liquid 
interactions in the study of the superconducting state follows 
the intuitively appealing approach of Ref.\ \protect\onlinecite{bet69},
which is consistent with 
more formal correlation function approaches.
\protect\cite{lar63,leg65} Rather than
starting with a band energy in the absence of electron-electron
interactions of $\epsilon^b_k = \hbar^2 k^2/(2m)$ as in Ref.\ 
\protect\onlinecite{bet69}, however, this article allows
$\epsilon^b_k$ to be an arbitrary function of ${\bf k}$ so as to be
able to account for anisotropic Fermi surface effects.  To 
form a Hamiltonian from $\epsilon^b_k$ the substitution
$\hbar {\bf k} \rightarrow -i\hbar \nabla - e{\bf A}/c$
is made. Also, this article studies only equilibrium properties,
and does not develop a kinetic equation.
Other studies of Fermi liquid
interactions in unconventional superconductivity include Refs.\ 
\protect\onlinecite{gro86,xu95}.  

The Hamiltonian describing the excitations of the superconducting 
state has the following form:
\begin{equation}
{\mathcal{H}} = \sum_k
\begin{array}{cc}
	[c_{k \uparrow}^\dagger & c_{-k \downarrow}]
\end{array}
\left[ \begin{array}{cc}
		\zeta_k +\lambda_k & \Delta_k \\
		\Delta_k & -\zeta_k + \lambda_k
	\end{array}	\right]
\left[ \begin{array}{c} 
		c_{k \uparrow} \\
		 c_{-k \downarrow}^\dagger 
	\end{array}	\right].
\label{H}
\end{equation}
Here $\zeta_k = \xi_k + \delta \varepsilon_k^{\mathcal{S}} + 
h_k^{\mathcal{S}}$ where $\xi_k$ is the Landau
quasiparticle energy
relative to the chemical potential (neglecting quasiparticle
interactions), 
$\lambda_k = \delta \varepsilon_k^{\mathcal{A}} + 
h_k^{\mathcal{A}}$, and $\Delta_k$ is the momentum-dependent
gap function appropriate for $d$-wave symmetry.  Fermi liquid 
interactions give a contribution to the energy of a quasiparticle 
with momentum k and spin $\sigma$ due to other excited quasiparticles
which is \protect\cite{bet69,pin89}. 
\begin{equation}
\delta \varepsilon_{k\sigma} =  \frac{1}{L^2} \sum_{k^\prime} 
	\left[ 	f^{\sigma \sigma}_{k k^\prime}
		\delta n_{k^\prime\sigma} 
	+ f^{\sigma \overline{\sigma}}_{k k^\prime} 
	\delta n_{k^\prime \overline{\sigma}}
	 \right].
\label{qpint}
\end{equation}
where $\overline{\sigma} \equiv - \sigma $, $n_{k\sigma} = 
\langle c_{k\sigma}^\dagger c_{c\sigma} \rangle$, and $\delta$ 
indicates a variation due to the excitation of other electrons 
and  holes, either by temperature or by the presence of external 
fields. (The factor $L^{-2}$ occurs in Eq.\ \ref{qpint} because the intention
is to develop a model applicable to superconductivity in a 
two-dimensional copper-oxide plane of a high $T_c$ superconductor
having area $L^2$.) An important step in the analysis, as described
qualitatively above, is the separation of 
the Fermi liquid interactions into ${\mathcal{S}}$ymmetric 
and ${\mathcal{A}}$ntisymmetric parts defined by
\begin{equation}
	\delta \varepsilon_k^{\mathcal{A}} = \frac{1}{2}
	\left[ 	\delta\varepsilon_{k\uparrow} - 	
	\delta\varepsilon_{-k\downarrow} \right],\ \ 
	\delta \varepsilon_k^{\mathcal{S}} 
	= \frac{1}{2}\left[ 	\delta\varepsilon_{k\uparrow} 
	+ \delta\varepsilon_{-k\downarrow} \right].
\label{saqpint}
\end{equation}

The quantities $h_k^{\mathcal{A}}$ and $h_k^{\mathcal{S}}$ in 
$\mathcal{H}$ represent
generalized external fields.  For example, in the case of an 
external magnetic field acting on the orbital motion of the electrons, 
the gap function will
acquire a complex phase.  This phase factor can be removed by a 
gauge transformation, $c_{k\sigma} \rightarrow c_{k\sigma} 
exp(i\theta)$, the end result of which is the addition of the field 
$h_k^{\mathcal{A}} =  {\bf v}^b_k \cdot {\bf p}_s,\ h_k^{\mathcal{S}} 
= 0$, where ${\bf v}^b_k \equiv \partial \epsilon^b_k/\partial {\bf k}$,
to the Hamiltonian.  
Here ${\bf p}_s = \hbar \nabla \theta - e{\bf A}/c$ is the 
superfluid momentum, which is assumed to be
sufficiently slowly varying spatially that its gradients 
can be neglected.  The
velocity  $ {\bf v}^b_k$ that appears in $h_k^{\mathcal{A}}$
is the bare band velocity, unrenormalized by the
electron-electron interaction, as noted following Eq. 12
of Ref.\ \protect\onlinecite{bet69}, and it is this same velocity that
appears in the expression for the quasiparticle contribution
to the current density (Eq.\ \ref{J} below).  On the other
hand, the electron-electron interaction contributes to
the quasiparticle energy $\xi_k$ defined in and following
Eq.\ \ref{H}, and hence affects the quasiparticle
velocity $v_F$ that occurs in $E^{(0)}_k$ below.  The differences
in these two velocities have important quantitative consequences for the
interpretation of the penetration depth data, as will be 
seen below.

The Hamiltonian of Eq.\ \ref{H} can be used to find the thermal
equilibrium expectation value of the electrical current density
operator giving, for
the electrical current density
 ${\bf J} = \eta_g{\bf p}_s + {\bf J}_{qp}$, with
the gauge contribution determined by
\begin{equation}
	\eta_g = \frac{e}{2L^2}\sum_{k\sigma}
	n_{k\sigma} \frac{1}{\hbar^2} \left(\frac{\partial^2}{\partial k_x^2} 
	+ \frac{\partial^2}{\partial k_y^2} \right)
	\epsilon^b_k,
	\label{etag}
\end{equation}
and the quasiparticle contribution  given by
\begin{equation}
	{\bf J}_{qp} =  \frac{e}{L^2} \sum_{k\sigma} 
	{\bf v}^b_k f(E_{k, \sigma}),
	\label{J}
\end{equation}
$f(E_{k, \sigma})$ is the Fermi-Dirac distribution function, 
and the $E_{k, \sigma}$ are the Bogoliubov quasiparticle energies defined 
below.  The case of an  external magnetic field H acting on the 
spin degrees of freedom is described by taking $h_k^{\mathcal{A}} 
= \mu_B H,\ h_k^{\mathcal{S}} = 0$. In both of these cases, the 
magnetic field acts only on the ${\mathcal{A}}$ntisymmetric 
mode, and
has no effect on the ${\mathcal{S}}$ymmetric mode of excitation.

In addition to causing changes in the 
energy of a quasiparticle (as in Eq.\ \ref{qpint}), excited quasiparticles
can give rise to changes in the gap function \protect\cite{bet69}.
There are however no changes that are linear in the superfluid momentum
\protect\cite{xu95}, and this 
effect  will therefore be neglected.

The diagonalization of the Hamiltonian of Eq.\ \ref{H}  gives
\begin{equation}
{\mathcal{H}} =
\sum_{k\sigma} E_{k, \sigma} \gamma_{k, \sigma}^\dagger 
	\gamma_{k, \sigma},\ \ 
	E_{k, \sigma} = E_{\sigma k} 
	+ \sigma(\delta \varepsilon_{\sigma k}^{\mathcal{A}} 
	+ h_{\sigma k}^{\mathcal{A}})
\label{Hdiagonal}
\end{equation}
where $\sigma = \pm 1$, $E_k = \sqrt{\zeta_k^2 + \Delta_k^2}$,
and the $\gamma_{k, \sigma}^\dagger$ are operators creating
Bogoliubov quasiparticles.

\begin{figure}%[b!] % fig 1
%\hspace{-.85in}
%\vspace{-2in}
\centerline{\epsfig{file=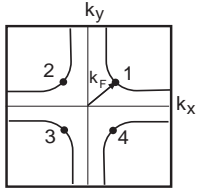,height=1in,width=1in}}
\vspace{10pt}
%\vspace{-2in}
\caption{ The labelling of the nodes on the Fermi surface
	of YBa$_2$Cu$_3$O$_{6+x}$}
\label{fig1}
\end{figure}

Later, the energy $E_k^{(0)} = \sqrt{\xi_k^2 + \Delta_k^2}$ 
describing the quasiparticle spectrum in the absence of other 
excited quasiparticles is also used.  
For a $d$-wave superconductor,
the quasiparticle energy can be parameterized 
\protect\cite{lee93} in the neighborhood of the
Fermi-surface nodal points (see Fig.\ \ref{fig1})
as $E_k^{(0)} 
= \sqrt{(p_1 v_F)^2 + (p_2 v_2)^2}$, where $p_1$ and 
$p_2$ are components of the momentum relative to the 
nodal point in directions perpendicular and parallel 
to the Fermi line.
At low temperatures, only quasiparticles close to these 
four points can be thermally excited. 

Using Eq.\ \ref{qpint} in Eq.\ \ref{saqpint}, keeping 
only terms up to linear order in the $\delta \varepsilon$'s 
and $h$'s, and
dropping some terms of order $k_B T/(\hbar k_F v_2)$
relative to those kept, yields 
the integral equations
\begin{equation}
	\delta \varepsilon_k^{\mathcal{S}}  =  \frac{1}{L^2}
	\sum_{k^\prime} f^{(+)}_{k k^\prime} 	
	    \left[ \frac{\xi_{k^\prime}}{E_{k^\prime}^{(0)}}
	f(E_{k^\prime}^{(0)}) 
	-\frac{\Delta_{k^\prime}^2}{E_{k^\prime}^{(0)3}} 
	(\delta \varepsilon_{k^\prime}^{\mathcal{S}} 
	+ h_{k^\prime}^{\mathcal{S}}) 	\right]
	\label{S}
\end{equation}
and
\begin{equation}
	\delta \varepsilon_k^{\mathcal{A}} = \frac{1}{L^2}\sum_{k^\prime} 
	f^{(-)}_{k k^\prime} \frac{\partial f}{\partial E_{k^\prime}^{(0)}}
	\left(\delta \varepsilon_{k^\prime}^{\mathcal{A}} 
	+ 	h_{k^\prime}^{\mathcal{A}} \right),	
	\label{A}
\end{equation}	
where $f^{(\pm)}_{k k^\prime} = f^{\sigma \sigma}_{k k^\prime}\pm
f^{\sigma \overline{\sigma}}_{k,-k^\prime}$.

Consider first  the 
${\mathcal{S}}$ymmetrical corrections to the quasiparticle energies
(Eq.\ \ref{S}), and
assume that there are no ${\mathcal{S}}$ymmetrical external
fields other than temperature, i.e. $h_k^{\mathcal{S}} = 0$ 
(as is the case for the external magnetic fields of most interest
in this article, which are purely ${\mathcal{A}}$ntisymmetrical). 
Then the only term driving a nonzero 
contribution to $\delta \varepsilon_k^{\mathcal{S}}$ is the term
on the right hand side 
proportional to $f(E_{k^\prime}^{(0)})$ and describing the
effect of temperature.
This term is proportional to
$T^3$, thus giving a $\delta \varepsilon_k^{\mathcal{S}} 
\propto T^3$, and will not be
important in contributing to the properties of interest at 
the temperatures satisfying $k_B T \ll \Delta_0$ ($\Delta_0$ 
is the maximum gap).  Thus $\delta \varepsilon_k^{\mathcal{S}}$ 
will be neglected in calculations below.

 Now from Eq.\ \ref{A}, which determines the 
${\mathcal{A}}$ntisymmetric corrections to the 
quasiparticle energies, it is clear that only the 
values of $\delta \varepsilon_k^{\mathcal{A}}$ and 
$h_k^{\mathcal{A}}$ at the Fermi surface nodes are 
relevant to the low energy properties.  Also, the 
solutions of Eq.\ \ref{A} can be classified according 
to the irreducible representation of the point group 
$C_{4v}$ (or 4mm) describing a tetragonal copper-oxide 
plane of a high T$_c$ superconductor, the independent 
solutions being
\begin{eqnarray}
	\delta \varepsilon^{\mathcal{A}}_{A_g} &=& 
	\left( \delta \varepsilon^{\mathcal{A}}_1 +
	\delta \varepsilon^{\mathcal{A}}_2 
	+ \delta \varepsilon^{\mathcal{A}}_3 
	+ \delta \varepsilon^{\mathcal{A}}_4 \right) /4 \nonumber \\
	\delta \varepsilon^{\mathcal{A}}_{xy} &=& 
	\left( \delta \varepsilon^{\mathcal{A}}_1 
	+ \delta \varepsilon^{\mathcal{A}}_2 
	- \delta \varepsilon^{\mathcal{A}}_3 
	- \delta \varepsilon^{\mathcal{A}}_4 \right) /4 \nonumber \\
	\delta \varepsilon^{\mathcal{A}}_{Ex} &=& 
	\left( \delta \varepsilon^{\mathcal{A}}_1 
	+ \delta \varepsilon^{\mathcal{A}}_2 
	+ \delta \varepsilon^{\mathcal{A}}_3 
	+ \delta \varepsilon^{\mathcal{A}}_4 \right) /4 \nonumber \\
	\delta \varepsilon^{\mathcal{A}}_{Ey} &=& 
	\left( \delta \varepsilon^{\mathcal{A}}_1
	+ \delta \varepsilon^{\mathcal{A}}_2 
	+ \delta \varepsilon^{\mathcal{A}}_3 
	+ \delta \varepsilon^{\mathcal{A}}_4 \right)/4 
	\label{irreps}
\end{eqnarray}
where the indices 1,2,3 and 4 refer to the four nodes in the 
excitation spectrum, as defined in Fig.\ \ref{fig1}.
The external 
fields $h^{\mathcal{A}}_k$ at the nodes can be similarly classified.

The solution of Eq.\ \ref{A} now yields
\begin{equation}
	\delta \varepsilon_{\Gamma}^{\mathcal{A}}(T) 
	= -h_{\Gamma}^{\mathcal{A}}F^{\mathcal{A}}_{\Gamma}(T)
	/[1 + F^{\mathcal{A}}_{\Gamma}(T)]
	\label{delta_e}
\end{equation}
with $F^{\mathcal{A}}_{\Gamma}(T) = f^{\mathcal{A}}_{\Gamma} 
ln(2) k_B T/(2\pi \hbar^2 v_F v_2)$. 
Here $\Gamma$ represents any of the irreducible representations present
in Eqs.\ \ref{irreps}.  The $f^{\mathcal{A}}_{\Gamma}$'s are defined by
\begin{eqnarray}
	f^{\mathcal{A}}_{A_g} 
		&=& f^a_{11} + f^a_{13} + 2 f^a_{12} \nonumber \\
	f^{\mathcal{A}}_{xy} 
		&=& f^a_{11} + f^a_{13} - 2 f^a_{12} \nonumber \\
	f^{\mathcal{A}}_{E} &=& f^s_{11} - f^s_{13}
	\label{fGamma}
\end{eqnarray}
where
\begin{equation}
	f^{s,a}_{k k^\prime} \equiv  f^{\sigma \sigma}_{k k^\prime} 
	\pm f^{\sigma \overline{\sigma}}_{k k^\prime}
	\label{fsa}
\end{equation}
are the symmetric and antisymmetric combinations of the 
Fermi-liquid parameters familiar from normal state 
analyses. \protect\cite{pin89}, and $f^a_{12}$ for example is
$f^a_{k k^\prime}$ for $k$ and $k^\prime$ at nodes 1 and 2,
respectively.

It is also useful to use Eq.\ \ref{S} to obtain an idea of how the 
${\mathcal{S}}$ymmetrical external 
fields are renormalized by Fermi-liquid interactions.   
It is clear from Eq.\ \ref{S}
that a knowledge of the Fermi-liquid interaction on the entire Fermi surface
is required and that $\delta \varepsilon^{\mathcal{S}}_k$ must be
determined on the entire Fermi surface.  To obtain a rough idea of
the nature of the solutions, consider a circular Fermi surface of
radius $k_F$ and look
for a solution of $A_g$ symmetry by considering a Fermi liquid interaction
$f^{(+)}_{k k\prime} = f^{\mathcal{S}}_{A_g}$, independent of $k$ and
$k^\prime$, and a ${\mathcal{S}}$ymmetrical external 
field $h^{\mathcal{S}}_{A_g}$ independent of $k$. The solution, which is
also independent of $k$ on the Fermi surface is
\begin{equation}
	\delta \varepsilon^{\mathcal{S}}_{A_g} = 
	-\frac{F^{\mathcal{S}}_{A_g}}{1+ 
	F^{\mathcal{S}}_{A_g}}h^{\mathcal{S}}_{A_g}
	\label{SAg}
\end{equation}
where $F^{\mathcal{S}}_{A_g} = f^{\mathcal{S}}_{A_g} k_F/(\pi \hbar v_F)$.
In contrast to the ${\mathcal{A}}$ntisymmetrical Fermi liquid parameters
$F^{\mathcal{A}}_{\Gamma}(T)$ obtained above, which go to zero linearly with
temperature in the superconducting state in the clean limit (and
hence have a dependence on temperature $T$ explicitly indicated), 
the ${\mathcal{S}}$ymmetrical Fermi liquid parameter
$F^{\mathcal{S}}_{A_g}$ is temperature independent and of approximately
the same magnitude as the corresponding normal state Fermi liquid
parameter.  The same can be seen to be true of the
${\mathcal{S}}$ymmetrical Fermi liquid parameters corresponding to
other irreducible representations of $C_{4v}$.  Note that the ratio
of the ${\mathcal{A}}$ntisymmetrical to the ${\mathcal{S}}$ymmetrical
Fermi-liquid $F$ parameters is 
$F^{\mathcal{A}}/F^{\mathcal{S}} \approx 
(f^{\mathcal{A}}/f^{\mathcal{S}})[k_B T/(\hbar k_F v_2)]$.

As noted above, the presence of a superfluid momentum contributes an
${\mathcal{A}}$ntisymmetrical external field to the Hamiltonian of
Eq.\ \ref{H}.  This external field corresponds to the E irreducible
representation of $C_{4v}$ with the $p_{sx}$ and $p_{sy}$ components of 
${\bf p}_s$ corresponding to the components $Ex$ and $Ey$ of 
Eq.\ \ref{irreps}.  The current density is thus easily evaluated using
Eq.\ \ref{J} with Eqs.\ \ref{Hdiagonal}, \ref{delta_e} and
\ref{fGamma}. The result is ${\bf J}_{qp} = \eta_{qp} {\bf p}_s$ 
where
\begin{equation}
	\eta_{qp}(T) =  -\frac{2 ln2 e (v^b_F)^2 k_B T}
	{[1 + F^{\mathcal{A}}_E(T)]\pi \hbar^2 v_F v_2}
	\label{rho}
\end{equation}
Note that the Fermi liquid correction does not 
alter the clean limit linear
in T contribution to the $\eta_{qp}(T)$,
but rather makes a $T^2$ contribution (using
$(1+F)^{-1} \approx (1 - F + ...)$).  Thus there are no Fermi liquid
corrections to the experimentally measured linear in $T$ contribution to
inverse square penetration depth. The penetration depth $\lambda$ is 
thus given by 
\begin{equation}
	\lambda^{-2}(T) = \lambda^{-2}(0) 
	- \frac{8 ln2 e^2}{c^2 \hbar^2} \alpha^2 \frac{v_F}{v_2} k_BT
	+ ...
 	\label{lambda}
\end{equation}
where $\alpha = (v^b_F/v_F)$, and $\lambda^{-2}(0)$ is determined by
$\eta_g$ given in Eq.\ \ref{etag}.  This has exactly the same form as Eq. 6 of
Ref.\ \protect\onlinecite{chi00}, which finds (from a detailed analysis of
a number of experiments) $\alpha^2 =$ 0.43 for Bi$_2$Sr$_2$CaCu$_2$O$_8$
and $\alpha^2 =$ 0.46 for YBa$_2$Cu$_3$O$_{7 - \delta}$.  The conclusions
here are however completely different from those drawn in Ref.\ 
\protect\onlinecite{chi00} which, based on previous theoretical work,
considered $\alpha^2$ to be a Fermi liquid correction with a not 
unreasonable value.  The conclusion of this article is that the 
experimentally determined value of $\alpha^2$ implies a value of $v^b_F$
significantly smaller than $v_F$, which is not physically
reasonable.  One of the essential features of strongly correlated
electron systems such as the copper-oxide superconductors is that
the strong electron-electron correlations are expected to produce
narrow energy bands and large quasiparticle masses, leading to
$v_F < v^b_F$.

The renormalization of the spin susceptibility due to Fermi-liquid 
interactions can be calculated in a similar way.  The
Zeeman interaction of the spin of an electron with the magnetic field
contributes an ${\mathcal{A}}$ntisymmetric external field of $A_g$
symmetry to the Hamiltonian.  It follows that the magnetic moment
per unit area of a copper oxide plane is
\begin{equation}
	M = -\frac{\mu_B}{L^2} \sum_k \left[f(E_{k,1}) - f(E_{k,-1}) \right]
	= \chi H
	\label{M}
\end{equation}
where 
\begin{equation}
	\chi (T) = \frac{\chi_0 (T)}{1 + F^{\mathcal{A}}_{A_g}(T)},\ \ 
	\chi_0 (T) = \frac{\mu_B^2 ln 2 k_B T}{\pi \hbar^2 v_F v_2}.
	\label{chi}
\end{equation}
Note that here also the low-temperature clean-limit linear in T magnetic
susceptibility is not changed by Fermi-liquid interactions. These
affect only terms of order $T^2$ and higher in the susceptibility.	  

%\begin{figure}%[b!] % fig 1
%\hspace{-.85in}
%\vspace{-2in}
%\centerline{\epsfig{file=walker_SBSIII_fig3.ps,height=4in,width=3.5in}}
%\vspace{10pt}
%\vspace{-2in}
%\caption{ The surface bound state contribution to the ZBCP for a (110) NIS 
%junction as predicted by Eqs.\ \ref{eq1} and \ref{eq2}.}
%\label{fig3}
%\end{figure}

This article has given a detailed description of both Fermi-liquid 
effects and band structure effects within the framework of a 
quasiparticle picture of the low-temperature properties of $d$-wave
superconductors.  This opens the way for a detailed quantitative 
experimental test of the quasiparticle picture of the low-temperature
properties. A classification of Fermi-liquid effects is given that
separates the strong renormalization and weak 
renormalization effects according to a 
symmetry property.
The application of the results to the interpretation of penetration
depth measurements is of particular interest.
This corresponds to the weak (and in fact negligible) Fermi-liquid
renormalization case,
and the ultimate conclusion 
is that the experimental results
imply a band
Fermi velocity which is smaller than the corresponding quasiparticle 
Fermi velocity, which is an unphysical result.
Clearly there are at present problems with the quantitative aspects 
of the quasiparticle picture of the low-temperature
properties of the high T$_c$ superconductors 
and further study is desirable.

I acknowledge stimulating discussion with L. Taillefer, the hospitality 
of P. Nozi\`{e}res, and the Theory Group of the
Institut Laue Langevin where much of this work was done,  and 
the support of the Canadian Institute for Advanced Research 
and of the Natural Sciences and Engineering Research Council of Canada.

\end{document}